\documentclass[%
reprint,
superscriptaddress,
showpacs,preprintnumbers,
 amsmath,amssymb,
 aps,
]{revtex4-1}

\usepackage{graphicx}
\usepackage{dcolumn}
\usepackage{bm}
\usepackage{color}

\begin{document}

\preprint{APS/123-QED}

\title{General atomistic approach for modeling metal-semiconductor interfaces using  density functional theory and non-equilibrium Green's function}

\author{Daniele Stradi}
\email{daniele.stradi@quantumwise.com}
 \affiliation{Center for Nanostructured Graphene (CNG), Department of Micro- and Nanotechnology (DTU Nanotech), {\O}rsteds Plads, Building 345B,
DK-2800 Kongens Lyngby, Denmark\\}
\affiliation{Quantumwise A/S, Fruebjergvej 3, Postbox 4, DK-2100 Copenhagen, Denmark}
\author{Umberto Martinez}
\affiliation{Quantumwise A/S, Fruebjergvej 3, Postbox 4, DK-2100 Copenhagen, Denmark}
\author{Anders Blom}
\affiliation{Quantumwise A/S, Fruebjergvej 3, Postbox 4, DK-2100 Copenhagen, Denmark}
\author{Mads Brandbyge}
 \affiliation{Center for Nanostructured Graphene (CNG), Department of Micro- and Nanotechnology (DTU Nanotech), {\O}rsteds Plads, Building 345B,
DK-2800 Kongens Lyngby, Denmark\\}
\author{Kurt Stokbro}
\affiliation{Quantumwise A/S, Fruebjergvej 3, Postbox 4, DK-2100 Copenhagen, Denmark}


\begin{abstract}
Metal-semiconductor contacts are a pillar of modern semiconductor
technology. Historically, their microscopic understanding has been
hampered by the inability of traditional analytical and numerical
methods to fully capture the complex physics governing their operating
principles. Here we introduce an atomistic approach based on density
functional theory and non-equilibrium Green's function, which includes
all the relevant ingredients required to model realistic
metal-semiconductor interfaces and allows for a direct comparison
between theory and experiments via $I$-$V_{bias}$ curves
simulations. We apply this method to characterize
 an Ag/Si interface relevant for
photovoltaic applications and study the rectifying-to-Ohmic transition as function of the semiconductor doping. We also demonstrate that the standard
``Activation Energy" method for the analysis of $I$-$V_{bias}$ data
might be inaccurate for non-ideal interfaces as it neglects
electron tunneling, and that finite-size atomistic models have problems in
describing these interfaces in the presence of doping, due to a poor
representation of space-charge effects. Conversely, the present method
deals effectively with both issues, thus representing a valid
alternative to conventional procedures for the accurate
characterization of metal-semiconductor interfaces.
\end{abstract}

\pacs{72.10.-d, 73.30.+y, 73.40.-c,71.15.Mb}

\maketitle

\section{\label{sec:Introduction}Introduction}

Metal-semiconductor (M-S) contacts play a pivotal role in almost any
semiconductor-based technology. They are an integral part of a broad range of devices with application as diverse as photovoltaics (PV) \cite{Saga2010}, transistors and diodes \citep{Sze,Kroemer2001}, and fuel-cells \cite{Carrette2001,Fu2015}.

The requirement of M-S interfaces with tailored characteristics, such as a specific resistance at the contact, has fueled research on the topic for decades \cite{Brillson1982,Tung2001}. Nevertheless, despite the high degree of sophistication of current semiconductor technology, the understanding of M-S interfaces at the microscopic level still constitutes a considerable challenge \cite{Ratner2008,Tung2014}. Even the structure of the interface itself, which is buried in the macroscopic bulk metal and semiconductor materials, represent a serious impediment, as it makes the direct exploration of the interface properties cumbersome.

A measure of the device current $I$ as a function of the applied bias $V_{bias}$ is a standard procedure to probe a M-S interface  \citep{Sze}, despite the drawback that the measured $I$-$V_{bias}$ curves do not provide any direct information on the interface itself, but rather on the full device characteristics. As such, it is common practice to interpret the $I$-$V_{bias}$ curves by fitting the data with analytical models, which are then used to extract the relevant interface parameters such as the Schottky barrier height $\Phi$  \cite{Schroder}.
As no general analytical model exists, the accuracy of this procedure critically depends on whether the model describes well the physical regime of the interface under scrutiny. Furthermore, most models disregard the atomistic aspect of the interface, although it is nowadays accepted that chemistry plays a dominant role in determining the electronic characteristics of the interface \cite{Tung1984,Schmitsdor1995,Tung2000,Tung2001,Tung2014}. These ambiguities complicate the assignment of the features observed in the measured spectra to specific characteristics of the M-S interface.

Conversely, atomistic electronic structure methods \cite{Martin} are an ideal tool for the characterization of M-S interfaces, and have been successfully employed over the years for their analysis \cite{Das1989,vanSchilfgaarde1990,Picozzi2000,Hoekstra1998,Tanaka2001,Ruini1997,Delaney2010,Hepplestone2014}. However, due to their computational cost, these studies have focussed on model interfaces described using finite-size models formed by few atomic layers  ({\em e.g.}, slabs), the validity of which is justified in terms of the local nature of the electronic perturbation due to the interface. For similar reasons, most studies have considered non-doped interfaces, as the models required to describe  a statistically meaningful distribution of dopants in the semiconductor would be excessively demanding \cite{Butler2012,Jiao2015}. Last but not least, these model calculations only describe the system at equilibrium ({\em i.e.}, at $V_{bias}$ = 0 V), thereby missing a direct connection with the $I$-$V_ {bias}$ measurements.

Here, we develop a general framework which attempts to overcome the limitations inherent in conventional electronic structure methods for simulating M-S interfaces. We employ density functional theory (DFT) \cite{Payne1992} together with the non-equilibrium Green's function (NEGF) method \cite{Brandbyge2002} to describe the infinite, non-periodic interface exactly. The DFT+NEGF scheme allows us to predict the  behavior of the M-S interface under working conditions by simulating the $I$-$V_{bias}$ characteristics of the interface at zero and at finite $V_{bias}$. To describe correctly the electronic structure of the doped semiconductor, we employ an exchange-correlation (xc) functional designed  {\em ad-hoc} to reproduce the experimental semiconductor band gap \cite{Tran2009}, and a novel spatially dependent effective scheme to account for the doping on the semiconductor side.

We apply this novel DFT+NEGF approach to study the characteristics of a Ag/Si interface relevant for PV applications  \cite{Kim1998,Weitering1993,Schmitsdor1995,Ballif2003,Li2009,Horteis2010,Garramone2010,
Hilali2005,Pysch2009,Li2010,Butler2011,Butler2012,Balsano2013}. Specifically, we focus on the (100)/(100) interface \cite{Kim1998,Butler2011} and on the dependence of $I$-$V_{bias}$ characteristics on the semiconductor doping -- notice that the method is completely general and can be used to describe other M-S interfaces with different crystalline orientations. We consider a range of doping densities for which the interface changes from rectifying to Ohmic. We demonstrate that the ``Activation Energy" (AE) method routinely employed to analyse M-S contacts systematically overestimate the value of $\Phi$, with an error that is both bias and doping dependent, due to the assumption of a purely thermionic transport mechanism across the barrier. Conversely, we show how an analysis of the $I$-$V_{bias}$ characteristics based on the DFT+NEGF electronic structure data provides a coherent picture of the rectifying-to-Ohmic transition as the doping is varied. Finally, we also show that a slab model does not provide a good representation of the interface electronic structure  when doping in the semiconductor is taken into account. This is due to the inability of the semiconductor side of the slab to screen the electric field resulting from the formation of the interface.

The paper is organized as follows. Section \ref{sec:Computational
  methods} and Section \ref{sec:System} describe the computational methods and the system models
employed in this work, respectively. Section \ref{sec:Device characteristics and
  validation of the activation energy model} presents the calculated
$I$-$V_{bias}$ characteristics and the validation of the AE method
based on the calculated data. Section \ref{sec:Electronic properties
  of the interface} deals with the analysis of the $I$-$V_{bias}$
curves in terms of the electronic structure of the interface as
obtained from the DFT+NEGF calculations. In Section
\ref{sec:Comparison of the two-probe with the slab model}, the
simulations are compared to finite-size slab models. The main conclusions are drawn in Section \ref{sec:Conclusions}.

\section{\label{sec:Computational methods}Computational methods}

The Ag(100)/Si(100) interface has been simulated using Kohn-Sham (KS) DFT as implemented in  \textsc{atomistix toolkit} \cite{ATK} (ATK). DFT \cite{Payne1992} and DFT+NEGF \cite{Brandbyge2002} simulations have been performed using a formalism based on a non-orthogonal pseudo-atomic orbitals \cite{Soler2002} (PAOs) basis set.

The one-electron KS valence orbitals are expanded using a linear
combination of double-$\zeta$ PAOs including polarization functions
(DZP). The confinement radii $r_c$ employed are 4.39 Bohr,  7.16 Bohr,
7.16 Bohr for the Ag 4$d$, 5$s$ and 5$p$ orbitals, and 5.40 Bohr,
6.83 Bohr, 6.83 Bohr for the Si 3$s$, 3$p$, 3$d$ orbitals, respectively. The ionic cores have been described using Troullier-Martins \cite{Troullier1991} norm-conserving pseudo-potentials \cite{Hamann1979}. The energy cutoff for the real-space grid used to evaluate the Hartree and xc contributions of the KS Hamiltonian has been set to 150 Ry. Monkhorst-Pack \cite{Monkhorst1976} grids of {\em k}-points have been used to sample the 3D (2D) Brillouin zone in the DFT (DFT+NEGF) simulations. We have used an $\mathrm{11\times11\times11}$ grid of {\em k}-points for the bulk calculations, and a {\em k}-points grid of $\mathrm{18\times9\times1}$ ($\mathrm{18\times9}$) for the DFT (DFT+NEGF)  simulations of the interface. Geometry optimizations have been performed by setting the convergence threshold for the forces of the moving atoms to 2 $\times$ 10$^{-2}$ eV/{\AA}. In all the simulations, periodic boundary conditions (PBCs) were used to describe the periodic structure extending along the directions parallel to the interface plane. In the slab DFT simulation, Dirichelet and Neumann boundary conditions were applied in the direction normal to the interface on the silver and silicon sides of the simulation cell, respectively, whereas in the DFT+NEGF simulations the same direction was described using Dirichelet boundary conditions at the two boundaries between the interface and the bulk-like electrodes.

\subsection{\label{sec:Spill-in terms}``Spill-in" terms}

\begin{figure}
    \includegraphics[width=0.4\textwidth]{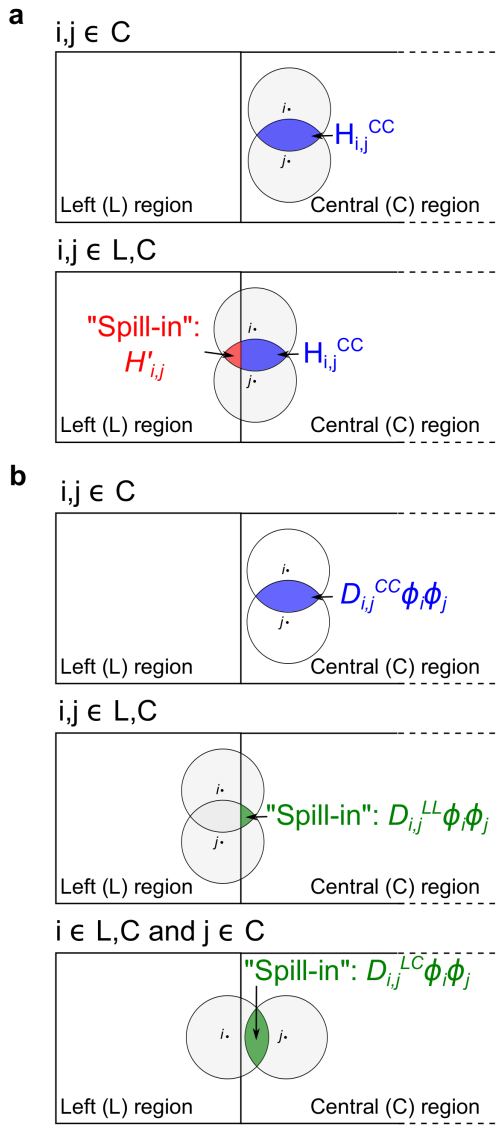}
    \caption{Scheme showing the two-center ``{\em spill-in}" terms used for the evaluation of the Hamiltonian (a) and the electronic density (b) of the C region for a pair of $s$ type PAOs $\phi_j$ and $\phi_i$ located close to the L/C boundary. In (a,b) the blue shaded regions indicate the integrals performed in the C region, whereas the red and green regions indicate the ``spill-in" terms for the Hamiltonian and for the electronic density, respectively.}
\label{fig:figure1}
\end{figure}

As described in Ref. \citenum{Brandbyge2002}, the DFT+NEGF method used to simulate the infinite, non-periodic Ag(100)/Si(100) interface relies on a two-probe setup, in which a left (L) and a right (R) semi-infinite electron reservoirs are connected through a central (C) region containing the interface. Once the chemical potentials $\mu_{L,R}$ of the reservoirs have been defined, a self-consistent (SCF) KS procedure is used to obtain the electronic density in the C region. The main quantity being evaluated in the SCF cycle is the density matrix required to express the electronic density of the C region in the basis of PAOs centered in the same region, $\bar{D}^{CC}$. Assuming $\mu_L > \mu_R$, $\bar{D}^{CC}$ takes the form

\begin{equation}
\begin{split}
\bar{D}^{CC} = & - \frac{1}{\pi}\int_{-\infty}^{\mu_R} \mathrm{Im}[\bar{G}^{CC}]d\epsilon \\
& - \frac{1}{\pi}\int_{\mu_R}^{\mu_L} \bar{G}^{CC}\mathrm{Im}[\bar{\Sigma}^{LL}] \bar{G}^{\dagger CC} d\epsilon,
\label{eq:negf_1}
\end{split}
\end{equation}

\noindent where $\bar{\Sigma}^{LL}$ is the self-energy matrix describing the coupling of the central region to the semi-infinite L reservoir, and the Green's function of the central region $\bar{G}^{CC}$ is obtained by

\begin{equation}
\bar{G}^{CC}(\epsilon) = [(\epsilon + i\delta)\bar{S}^{CC}-\bar{H}^{CC}-\bar{\Sigma}^{LL}-\bar{\Sigma}^{RR}]^{-1},
\label{eq:negf_2}
\end{equation}

\noindent with $\bar{S}^{CC}$ and $\bar{H}^{CC}$ being the overlap and Hamiltonian matrices associated with the PAOs centered at the C region, and $\bar{\Sigma}^{RR}$ being the self-energy matrix of the R reservoir.

However, even if the DFT+NEGF method provides an elegant scheme to evaluate $\bar{D}^{CC}$, it should be noticed that solving Eqs. \ref{eq:negf_1}-\ref{eq:negf_2} is not sufficient to obtain the correct Hamiltonian and the electronic density of the C region. The reason is that the relevant integrals involved in  Eqs.  \ref{eq:negf_1}-\ref{eq:negf_2} are evaluated only in the region of space encompassing the C region, and only for the atoms localized in that region. As a consequence, the tails of the PAOs located close to both sides of the L/C and R/C boundaries, which penetrate into the neighboring regions, are not accounted for (see Fig. \ref{fig:figure1}). To correct this behavior, we introduce additional corrective terms, that we name ``spill-in". For the Hamiltonian, corrective terms are applied to both the two-center and three-center integrals. Specifically, if two PAOs $\phi_i$ and $\phi_j$ centered in the C region lie close to a boundary, {\em e.g.} the L/C one, the corrected Hamiltonian in Eq. \ref{eq:negf_2} will include also the  matrix element $H_{i,j}^{\prime} = \langle  \phi_i | V_{eff}^{LL}(\mathbf{r}) | \phi_j \rangle$ associated with the tail of the PAOs protruding into the L region, $V_{eff}^{LL}(\mathbf{r})$ being the periodic KS potential of the semi-infinite L reservoir (Fig. \ref{fig:figure1}a). Similar arguments hold also for the Hamiltonian three-center non-local terms.  For the electronic density of the C region, additional contributions are included for each pair of PAOs $\phi_i$ and $\phi_j$ located close to a boundary when at least one of them is centered at the neighboring reservoir region. In total, two new contributions must be added to the electronic density evaluated using Eqs. \ref{eq:negf_1}-\ref{eq:negf_2} for each pair of PAOs at each boundary. For the L/C boundary, these are (Fig. \ref{fig:figure1}b):

\begin{equation}
\begin{split}
n^{LL}  & =  \sum_{i,j} D_{i,j}^{LL} \phi_{i}^L \phi_{j}^L, \\
n^{LC} & =   \sum_{i,j} D_{i,j}^{LC} \phi_{i}^L \phi_{j}^C,
\end{split}
\end{equation}

\noindent which can be further distinguished based on whether both ($\bar{D}^{LL}$) or just one ($\bar{D}^{LC}$) of the two PAOs involved is centered at the L region. In the calculations presented in this work, the ``spill-in" terms are independent of the applied bias voltage. This is justified as we checked that the non-periodic KS potential at the boundary of the C region for each value of the applied bias matches smoothly with the periodic KS potential of the neighboring reservoir, {\em i.e.} that the ``screening approximation" is verified -- see Ref. \citenum{Brandbyge2002} for additional details. We stress that including these ``spill-in" terms is essential to ensure a stable and well-behaved convergence behavior of the SCF cycle, which turns out to be especially  important for heterogeneous systems such as the Ag(100)/Si(100) interface investigated in this work.

\begin{figure}
    \includegraphics[width=0.35\textwidth]{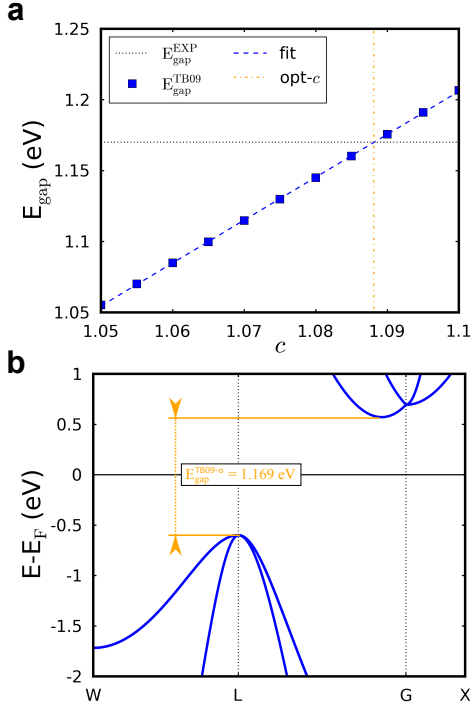}
    \caption{(a) Fitting procedure for the TB09 xc-functional $c$ parameter. Squares (blue): calculated indirect band gap of bulk silicon $\mathrm{E_{gap}^{TB09}}$ obtained for different values of the $c$ parameter. Dashed line (blue): fit to the computed data of $\mathrm{E_{gap}^{TB09}}$ {\em vs.} $c$ obtained by linear regression. Dotted line (black): experimentally measured bulk silicon band gap \cite{Kittel}. Dashed-dotted line (orange): optimal value of the $c$ parameter (opt-$c$), obtained as the intersect between $\mathrm{E_{gap}^{TB09}}$ and the fit to the $\mathrm{E_{gap}^{TB09}}$ data. (b) Region around the indirect band gap in the bulk silicon band structure calculated using the optimal $c$ parameter determined from (a).}
\label{fig:figure2}
\end{figure}

\subsection{\label{sec:Exchange-correlation potential}Exchange-correlation potential}

Further complications in describing the Ag(100)/Si(100) interface arise from the fact that one of its sides is semiconducting. In fact, a major problem affecting the description of metal-semiconductor interfaces is the severe underestimation of the semiconductor band gap in DFT calculations using (semi-)local xc-functionals based on the local density approximation (LDA) or on the generalized gradient approximation (GGA) \cite{Choen2012}. For model calculations based on few-layer thick fully periodic systems, such an underestimation has been shown to result in unrealistically low Schottky barriers at the interface \cite{Das1989,Godby1990}. In order to remedy this drawback, we have  evaluated the electronic structure of the LDA-optimized interface geometries using the Tran-Blaha meta-GGA xc-functional (TB09) \cite{Tran2009}. The TB09 xc-functional has been shown to provide band gaps in excellent agreement with the experiments for a wide range of semiconductors including silicon, at a computational cost comparable to that of conventional (semi-)local functionals. In the TB09 xc-functional, the exchange potential $\upsilon^\mathrm{TB09}_x(\mathbf{r})$ depends explicitly on electron kinetic energy $\tau(\mathbf{r})$,

\begin{equation}
\upsilon^\mathrm{TB09}_x(\mathbf{r}) =
c\upsilon^\mathrm{BR}_x(\mathbf{r}) +
\frac{3c-2}{\pi}
\sqrt{\frac{4\tau(\mathbf{r})}{6\rho(\mathbf{r})}},
\label{eq:tb09-1}
\end{equation}

\noindent with $\tau(\mathbf{r})=1/2\sum_{i=1}^N |\nabla\psi_i(\mathbf{r})|^2$, $N$ being the total number of KS orbitals, $\psi_i(\mathbf{r})$ the $i$-th orbital,  $\rho(\mathbf{r})$ the electronic density and $\upsilon^\mathrm{BR}_x(\mathbf{r})$ the Becke-Roussel exchange potential \cite{Becke1989}. The parameter $c$ in equation (\ref{eq:tb09-1}) is evaluated self-consistently and takes the form

\begin{equation}
 c = \alpha + \beta \left[ \frac{1}{\Omega} \int_\infty \frac{|\nabla\rho(\mathbf{r})|}{\rho(\mathbf{r})} d\mathbf{r} \right]^{\frac{1}{2}},
 \label{eq:tb09-2}
\end{equation}

\noindent where $\Omega$ is the volume of the simulation cell and the two empirical parameters $\alpha = -0.012$ (dimensionless) and $\beta = 1.023$ Bohr$^{\frac{1}{2}}$ have been fitted to reproduce the experimental band gaps of a large set of semiconductors \cite{Tran2009}. To obtain a description as accurate as possible of the semiconductor band gap at the Si(100) side of the interface, we have tuned the value of the $c$ parameter in order to reproduce the experimentally measured band gap of bulk silicon, $E_\mathrm{gap}^\mathrm{exp}$ = 1.17 eV \cite{Kittel}. This has been accomplished by calculating the band gap of bulk silicon at fixed values of the $c$ parameter in a range around the self-consistently computed value in which the variation of $E_\mathrm{gap}^\mathrm{TB09}$ with $c$ is linear. Then, the optimal value of $c$ has been determined as the intersect between the value of $E_\mathrm{gap}^\mathrm{exp}$ and a linear fit to the computed values of $E_\mathrm{gap}^\mathrm{TB09}$ (Fig. \ref{fig:figure2}a). Using the TB09 xc-functional with the $c$ parameter fixed at the optimal value determined using this procedure (hereafter, TB09-o), we calculate the indirect band gap of bulk silicon to be $E_\mathrm{gap}^{\mathrm{TB09-o}}$ = 1.169 eV (Fig. \ref{fig:figure2}b), in excellent agreement with the value 1.17 eV. The TB09-o functional has been used for all the electronic structure and transport analyses of the Ag(100)/Si(100) interface reported in this work. We have checked that the band structure of bulk silver calculated using the TB09-o functional is very similar to that calculated using the LDA, which is known to perform well for noble metals.

\subsection{\label{sec:Semiconductor doping}Semiconductor doping}

\begin{figure*}
    \includegraphics[width=0.8\textwidth]{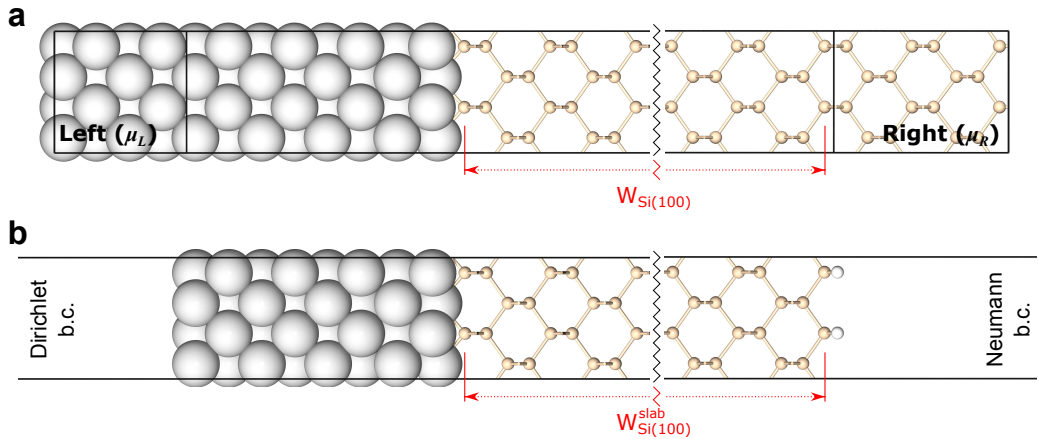}
    \caption{Geometries employed to simulate the Ag(100)/Si(100) interface using two-probe models (a) or slab models (b). Silver, silicon and hydrogen atoms are shown in grey, beige and white, respectively.}
\label{fig:figure3}
\end{figure*}

The last requirement to describe realistically the electronic structure of the Si(100)/Ag(100) interface is to account for the doping on the silicon side of the interface. Here, doping is achieved in an effective scheme by introducing localized charges bound to the individual silicon atoms. More specifically, in ATK \cite{ATK} the total self-consistent electronic density $\rho_\mathrm{tot}(\mathbf{r})$ is defined as \cite{Soler2002}:

\begin{equation}
\rho_\mathrm{tot}(\mathbf{r}) = \delta \rho(\mathbf{r}) + \sum_I^\mathrm{N_{atoms}} \rho_{I}(\mathbf{r}),
\label{eq:compensation_charge-1}
\end{equation}

\noindent where $\sum_I^\mathrm{N_{atoms}} \rho_{I}(\mathbf{r})$ is the sum of the atomic densities of the individual neutral atoms of the system. As each atomic density $\rho_{I}(\mathbf{r})$ is a constant term, it can be  augmented with a localized ``compensation" charge having the opposite sign of the desired doping density, which acts as a carrier attractor by modifying the electrostatic potential on the atom. Using these ``compensation" charges, an effective doping can be achieved both in the DFT and in the DFT+NEGF simulations. In the former, the ``compensation" charge added to each silicon atom is neutralized by explicitly adding a valence charge of the opposite sign, so that the system remains charge neutral. In the latter, the ``compensation" charge is neutralized implicitly by the carriers provided by the reservoirs, and the system is maintained charge neutral under the condition that the intrinsic electric field in the system is zero. This effective doping scheme has the advantage of (i) not depending on the precise atomistic details of the doping impurities, and (ii) being completely independent of the size and exact geometry of the system.

\section{\label{sec:System}System}

In order to obtain a reliable description of the Ag(100)/Si(100) interface, we have followed a stepwise procedure. Initially, we have carried out a preliminary screening of the interface geometries and bonding configurations by using a 2$\times$1 slab model formed by a 6-layers Ag(100) slab interfaced with a 9-layers unreconstructed Si(100) slab. The calculated bulk lattice constants of silicon ($a_\mathrm{Si}$ = 5.41 \AA) and silver ($a_\mathrm{Ag}$ = 4.15\AA) are in good agreement with those reported in the literature \cite{Butler2012}. To match the Ag(100) and the Si(100) surface, we have applied an isotropic compressive strain $\epsilon_{xx}$ = $\epsilon_{yy}$ = --0.0793 along the surface lattice vectors $\mathbf{v}_{1,2}$ of the Ag(100) surface. We have checked that in the compressed Ag(100) structure, the dispersion of the $s$-band and its position with respect to the $d$-band are very similar to those calculated using the equilibrium value of $a_\mathrm{Ag}$. The Si(100) surface opposite to the interface has been passivated with hydrogen atoms. The geometry of the resulting 15-layers slab has then been optimized using the LDA by keeping the farthest (with respect to the interface plane) 4 layers of the Ag(100) surface frozen, and by allowing the farthest (with respect to the interface plane) 4 layers of the Si(100) slab to move as a rigid body, thereby freezing only the interatomic distances and angles. All the remaining atoms have been allowed to fully relax. Different starting guesses for the interface structure have been tested, corresponding to different configurations of the Si(100) dangling bonds with respect to the high symmetry {\em fcc} sites of the Ag(100) surface. The lowest energy configuration among those considered, corresponding to the Si(100) dangling bonds sitting above the ``hollow" {\em fcc} sites of the Ag(100) surface, has then been used as a representative model of the interface.

Starting from the lowest energy configuration obtained using the
15-layers slab, we have then constructed more realistic models of the
interface. Specifically, we have expanded the bulk regions of the
15-layer slab to create two-probe setups effectively describing the
infinite, non-periodic interface (Fig. \ref{fig:figure3}a). A final
geometry optimization has been carried out using a two-probe setup in
which the C region has been described by 8 Ag(100) layers and an
undoped silicon layer having a total width $W_\mathrm{Si(100)}^\mathrm{CC}$ =
47.84 \AA. The optimized geometry has been used to construct two-probe
setups in which the doping of the silicon side has been taken into
account using the effective doping method described in Section
\ref{sec:Semiconductor doping}. We have considered doping densities of
    {\em n}-type carriers ($n_\mathrm{d}$) in the experimentally
    relevant range [10$^{18}$ cm$^{-3}$ -- 10$^{20}$ cm$^{-3}$]. As
    discussed in more detail in Section \ref{sec:Results}, the width
    of the Si(100) layer needed to describe accurately the interface
    in the two-probe simulations depends on the size of the
    depletion region ($W_\mathrm{D}$) on the silicon side of the
    interface. The relation between $W_\mathrm{D}$ and $n_\mathrm{d}$
    is 1/$W_\mathrm{D}$ $\propto$  n$_\mathrm{d}^{1/2}$, so that
    progressively narrower C regions can be used as the doping level is increased without any loss in accuracy. Therefore, in the following, the results presented for $n_\mathrm{d}$ =  10$^{20}$ cm$^{-3}$, $n_\mathrm{d}$ =  10$^{19}$ cm$^{-3}$ and $n_\mathrm{d}$ = 10$^{18}$ cm$^{-3}$ refer to calculations performed with C regions of widths $W_\mathrm{Si(100)}^\mathrm{CC}$ = 47.84 \AA, $W_\mathrm{Si(100)}^\mathrm{CC}$ = 197.436 \AA\ and $W_\mathrm{Si(100)}^\mathrm{CC}$ = 447.92 \AA, respectively. We have checked that reducing the width of the C region does not have any effect on the results, as long as all the space-charge effects due to the presence of the interface take place within the screening region. Furthermore, we notice how using a two-probe setup also allows to simulate the characteristics of the interface when the L and R reservoirs are set at two different chemical potentials $\mu_L \neq \mu_R$ due to an applied bias voltage $qV_{bias} = \mu_R - \mu_L$. As will become clear later, this allows for a direct comparison to experiments and for  analyzing the electronic structure of the interface under working conditions.

Finally, to understand to which extent the slab model is able to describe accurately the electronic structure of the infinite, non-periodic interface, we have also considered slab models having a similar interface geometry as that used in the two probe setup (Fig. \ref{fig:figure3}b). Both short and long slab models have been constructed, in which the width of the Si(100) layer used to describe the silicon side of the interface has been set to either $W_\mathrm{Si(100)}^\mathrm{slab(short)}$ = 38.33 \AA\, or $W_\mathrm{Si(100)}^\mathrm{slab(long)}$ = 98.62 \AA. Notice how these values of $W_\mathrm{Si(100)}^\mathrm{slab}$ are many times larger than those used for similar studies of the Ag(100)/Si(100) interface reported in the literature \cite{Butler2011}.

\begin{figure}
    \includegraphics[width=0.4\textwidth]{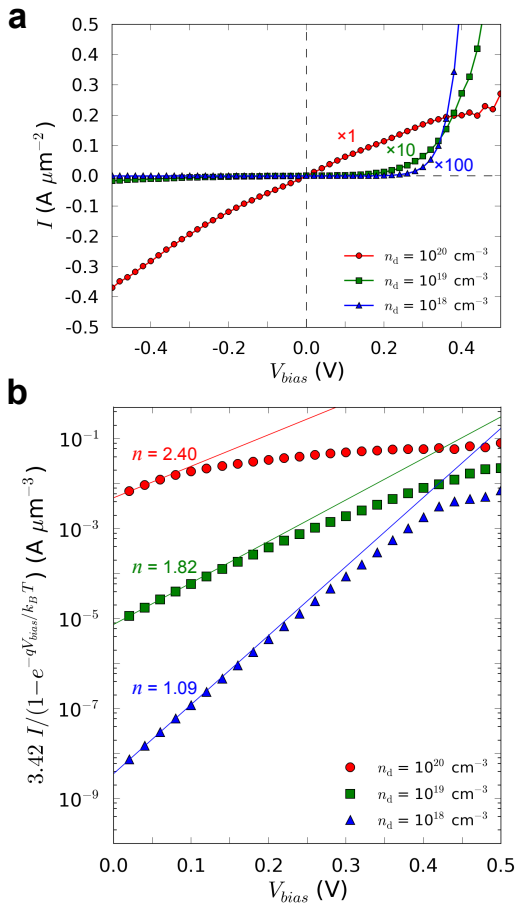}
    \caption{Calculated $I$-$V_\mathrm{bias}$ (a) and forward bias $I/(1-e^{q|V_{bias}|/k_BT})$-$V_{bias}$ (b) characteristics at $n_\mathrm{d}$ = 10$^{18}$ cm$^{-3}$ (blue triangles), $n_\mathrm{d}$ = 10$^{19}$ cm$^{-3}$ (green squares), $n_\mathrm{d}$ = 10$^{20}$ cm$^{-3}$ (red dots). In (a), the values of $I$ at $n_\mathrm{d}$ = 10$^{18}$ cm$^{-3}$ and $n_\mathrm{d}$ = 10$^{19}$ cm$^{-3}$ have been multiplied by a factor 10 and 100, respectively. The solid lines in (b) are fit to the data in the range 0.02 $\leq$ $V_\mathrm{bias}$ $\leq$ 0.08 V using Eq. \ref{eq:thermionic_1}. The ideality factor {\em n} extracted from the slope of each fitted curve is reported using the same color as the corresponding curve.}
\label{fig:figure5}
\end{figure}

\section{\label{sec:Results}Results}

\subsection{\label{sec:Device characteristics and validation of the activation energy model}Device characteristics and validation of the activation energy model}

Fig. \ref{fig:figure5}a shows the current--voltage ($I$-$V_{bias}$) characteristics calculated for the two-probe setup at low ($n_\mathrm{d}$ = 10$^{18}$ cm$^{-3}$), intermediate ($n_\mathrm{d}$ = 10$^{19}$ cm$^{-3}$) and high ($n_\mathrm{d}$ = 10$^{20}$ cm$^{-3}$) doping densities of the Si(100) side of the interface. A strong dependence on the doping concentration is evident. At low doping, the interface shows a well-defined Schottky diode-like behavior: the forward bias ($V_{bias}$ $>$ 0 V) current increases about six orders of magnitude in the range of $V_{bias}$ [0.02 V -- 0.5 V], whereas the reverse bias one (V$_\mathrm{bias}$ $<$ 0 V) varies only within one order of magnitude in the corresponding range. The diode-like asymmetry in the $I$-$V_{bias}$ curves persists at intermediate doping, although it is less pronounced than at low doping; the current at forward bias and reverse bias varying within three and two orders of magnitude, respectively. The scenario changes qualitatively at high doping as the $I$-$V_{bias}$ curve becomes highly symmetric, suggesting an Ohmic behavior of the interface.

\begin{figure*}
    \includegraphics[width=0.9\textwidth]{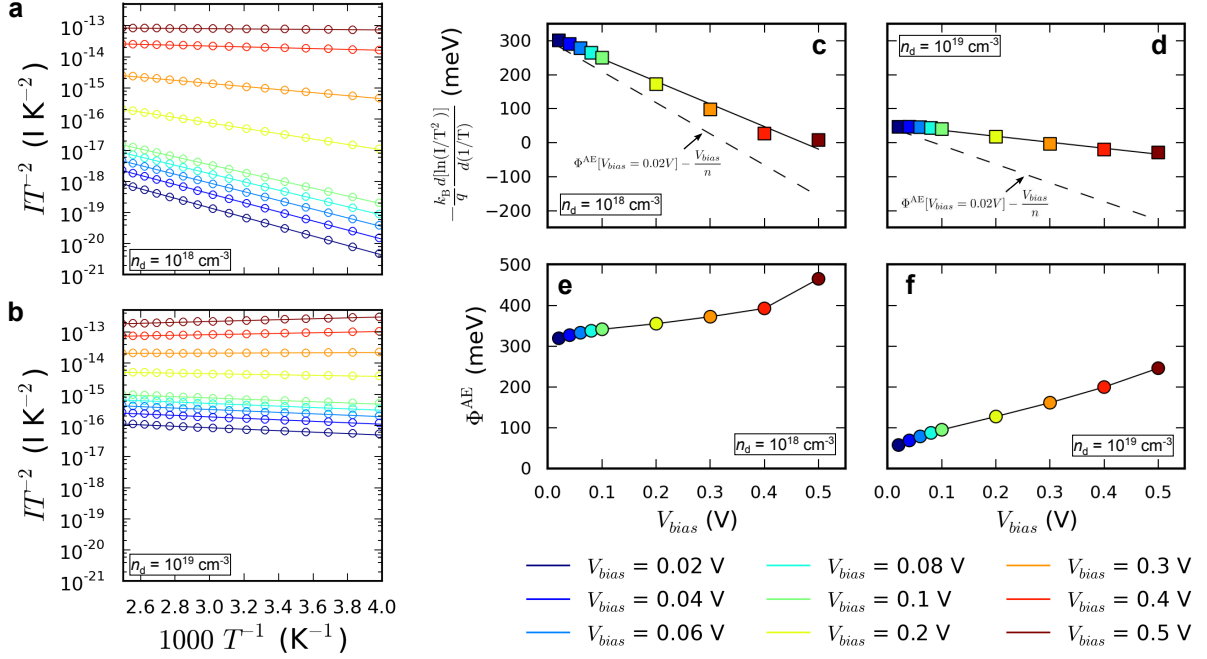}
    \caption{(a,b) Empty dots: calculated $I$-$T$ data at different bias voltages for $n_\mathrm{d}$ = 10$^{18}$ cm$^{-3}$ (a) and $n_\mathrm{d}$ = 10$^{19}$ cm$^{-3}$ (b). Solid lines: fit to the simulated data using Eq. \ref{eq:thermionic_3}. (c,d) Left-hand side (filled dots) and right-hand side (dashed line) of Eq. \ref{eq:thermionic_3} as a function of $V_{bias}$. The values of the left-hand side have been extracted from the slope of the fitted $I$-$T$ curves in (a,b). The solid lines are linear fits to the data. The right-hand side of Eq. \ref{eq:thermionic_3} has been plotted using the value of $\Phi^\mathrm{AE}$ calculated at $V_{bias}$ = 0.02 V, which approaches the value of $\Phi$ at $V_{bias}$ = 0 V. (e,f) Schottky barrier height $\Phi^\mathrm{AE}$ evaluated using Eq. \ref{eq:thermionic_3} as a function of $V_{bias}$.}
\label{fig:figure6}
\end{figure*}

According to thermionic emission theory, the $I$-$V_{bias}$ characteristics of a Schottky diode can be described by \citep{Sze}

\begin{equation}
I = I_0\, \Big[ e^{\frac{q V_{bias}}{n k_B T}}-1 \Big],
\label{eq:thermionic_1}
\end{equation}

\noindent where $q$ is the elementary charge, $k_B$ is the Boltzmann constant, $T$ is the temperature, $I_0$ is the saturation current and $n$ is the so-called ideality factor. The latter accounts for the deviation of the $I$-$V_{bias}$ characteristics from those of an ideal diode, for which $n$ = 1. Fitting the simulated data at forward bias to Eq. \ref{eq:thermionic_1} allows to extract $n$ from the slope of the fitted curves. In Fig. \ref{fig:figure5}b the fitted curves are compared to the forward bias data. The latter are presented using an alternative form of Eq. \ref{eq:thermionic_1},

\begin{equation}
{I} = I_0\, e^{\frac{q V_{bias}}{n k_B T}}\, \Big(1 - e^{-\frac{q V_{bias}}{k_B T}}\Big),
\label{eq:thermionic_1b}
\end{equation}

\noindent which allows for a better comparison with the fitted curves as $I/(1-e^{-qV_{bias}/k_BT})$ varies exponentially with $V_{bias}$ in the fitting interval considered, {\em viz.} 0.02 V $\leq$ V$_{bias}$ $\leq$ 0.08 V.

At low doping, $n$ = 1.09, indicating that the system behaves essentially as an ideal Schottky diode. At intermediate doping, $n$ = 1.82, and the system deviates significantly from the ideal behavior. At high doping, $n$ = 2.40, consistently with the observation that the system does not behave anymore as a Schottky diode.

The $I$--$V_{bias}$ simulation allows to test the reliability of the experimental procedures used to extract the Schottky barrier $\Phi$. In particular, we focus on the so-called ``Activation-Energy" (AE) method, which does not require any {\em a priori} assumption on the electrically active interface area $A$ \cite{Sze}. In the AE method the $I$--$T$ dependence is measured at a small constant $V_{bias}$. Over a limited range of $T$ around room temperature, assuming that the Richardson constant $A^{*}$ and $\Phi$ are constant, the $I$-$T$ characteristics can be described by the expression

\begin{equation}
I T^{-2} = AA^{*}\, e^{-\frac{q\Phi^{AE}}{k_B T}}\, e^{\frac{q(V_{bias}/n)}{k_B T}}.
\label{eq:thermionic_2}
\end{equation}

Following Eq. \ref{eq:thermionic_2}, the Schottky barrier height $\Phi^\mathrm{AE}$ can be extracted from the  $\ln(I/T^2)$ {\em vs.} $1/T$ data using

\begin{equation}
- \frac{k_B}{q} \frac{d[\ln(I/T^2)]}{d(1/T))} = \Phi^\mathrm{AE} -\frac{V_{bias}}{n},
\label{eq:thermionic_3}
\end{equation}

\noindent in which $n$ is the ideality factor extracted above.

\begin{figure*}
    \includegraphics[width=0.8\textwidth]{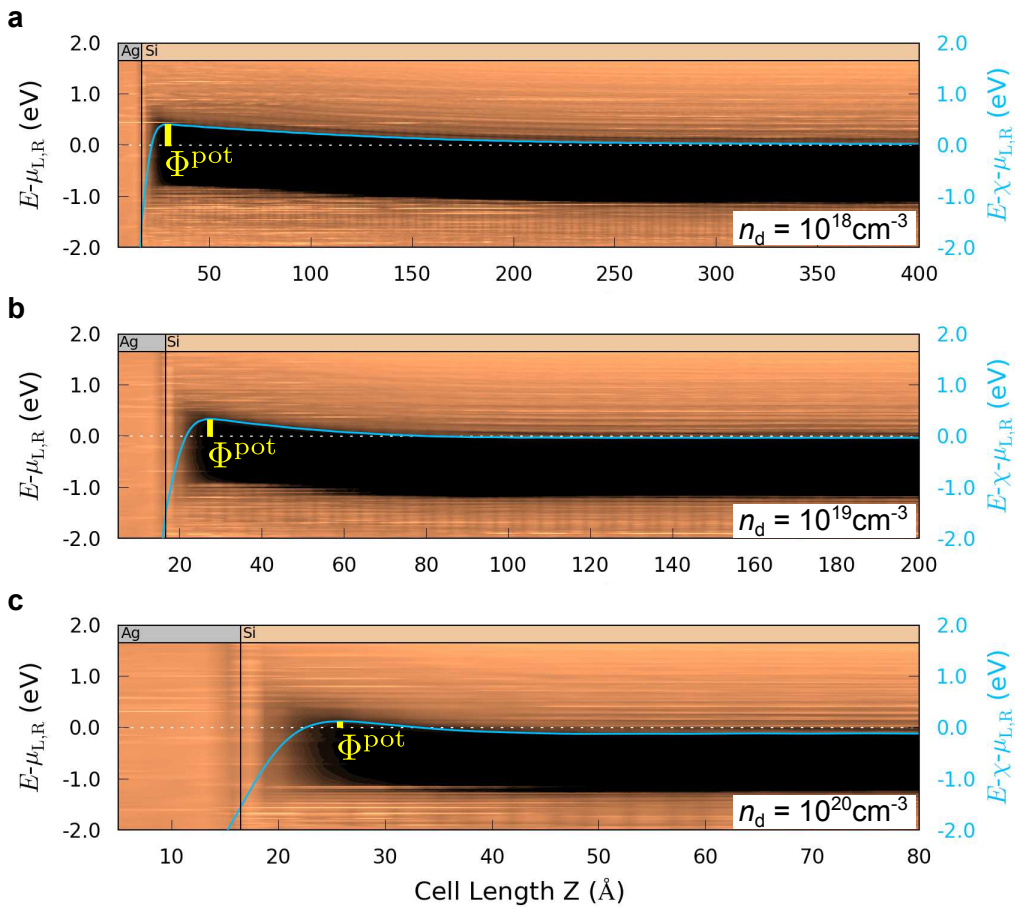}
    \caption{Local density of states (LDOS) of the two-probe setup at equilibrium for $n_\mathrm{d}$ =  10$^{18}$ cm$^{-3}$ (a), $n_\mathrm{d}$ =  10$^{19}$ cm$^{-3}$ (b) and $n_\mathrm{d}$ =  10$^{20}$ cm$^{-3}$ (c). The energy on the vertical is relative to the system chemical potential $\mu_\mathrm{L,R}$. Regions of low (high) LDOS are shown in dark (bright) color. The blue line in each panel indicates the macroscopic average of the Hartree potential $\langle V_H \rangle$ subtracted the electron affinity of bulk Si and  $\mu_\mathrm{L,R}$. The yellow vertical line in each panel indicates the associated $\Phi^\mathrm{pot}$.}
\label{fig:figure4}
\end{figure*}

Fig. \ref{fig:figure6}a,b shows the simulated AE plots (as Arrhenius plots) at different values of $V_{bias}$ for low and intermediate doping densities, at which the interface still displays clear Schottky diode--like characteristics. The $I$-$T$ dependence has been evaluated in a linear response fashion, using the Landauer-B\"uttiker expression for the current, $I = \frac{2q}{h}\int T(E,\mu_L,\mu_R) [f(\frac{E-\mu_L}{k_B T}) - f(\frac{E-\mu_R}{k_B T})] dE$ with the transmission coefficient $T(E,\mu_L,\mu_R)$ evaluated self-consistently at an electron temperature of 300 K. Fully self-consistent simulations performed for selected temperatures show that this approach is valid within the range of $T$ considered, 250 K $\leq T \leq$ 400 K.

Ideally, for a given doping the Schottky barrier depends exclusively
on the M-S energy level alignment at the interface and
therefore, disregarding image-force lowering effects, should remain
constant with $V_{bias}$ \citep{Sze}. This implies that in
Eq. \ref{eq:thermionic_3}, the left-hand side should equal the
right-hand side at any value of $V_{bias}$. However, in the present
case this condition is not verified, as the variation of the
right-hand side term with $V_{bias}$ is larger than that of the
left-hand side term (see Fig. \ref{fig:figure6}c,d). Indeed, for
$n_\mathrm{d}$ =  10$^{18}$ cm$^{-3}$ ($n_\mathrm{d}$ =  10$^{19}$
cm$^{-3}$), a linear fit to the calculated values of the left-hand
side of Eq. \ref{eq:thermionic_3} gives a slope of --664 meV/V (--177
meV/V), whereas the slope associated to the variation of the
right-hand side term is --917 meV/V (--549 meV/V).

Following the procedure in the AE method we use the
value of $n$ obtained from Fig.~\ref{fig:figure5} to subtract the
bias dependence. The result is shown in Fig.
\ref{fig:figure6}e,f and it can be seen that this leads to an
unphysical increase of $\Phi^\mathrm{AE}$ with $V_{bias}$. The error becomes more severe as $n_\mathrm{d}$
is increased. At low (intermediate) doping, $\Phi^\mathrm{AE}$ varies
from by 30$\%$ (325$\%$) in the range of V$_{bias}$ considered,
leading to a change $\Delta\Phi^\mathrm{AE}$ = 3.73 $k_B T$
($\Delta\Phi^\mathrm{AE}$ = 7.31 $k_B T$). Thus, the intrinsic
accuracy of the AE method depends strongly on multiple factors. On the
one side, the non-linear increase in $\Phi^\mathrm{AE}$ with
$V_{bias}$ suggests that using a single value of $V_{bias}$ is not
sufficient to obtain an accurate estimate of $\Phi$. On the other
side, the change in $\Delta\Phi^\mathrm{AE}$ with $n_\mathrm{d}$ at a
given $V_{bias}$ indicates that the AE method is unsuited for
comparative analyses of the variation of $\Phi$ with doping. These
facts call for a more direct and general strategy for the
characterization of M-S interfaces under working
conditions.

\subsection{\label{sec:Electronic properties of the interface}Electronic properties of the interface}

\begin{figure}
    \includegraphics[width=0.5\textwidth]{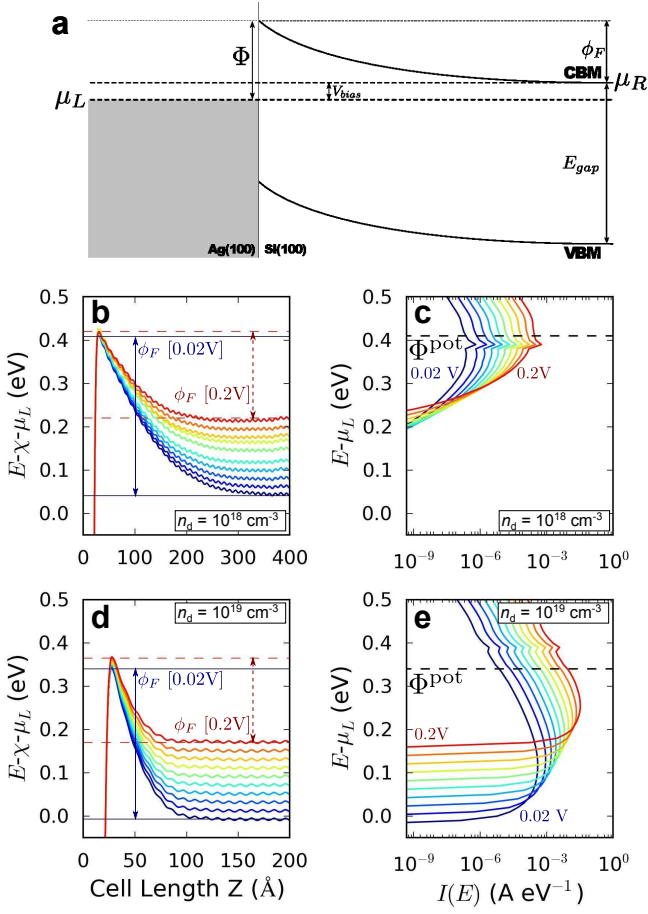}
    \caption{(a) Scheme of the electronic structure of the Ag/Si interface at forward bias voltage. (b) Profile of $\langle V_H \rangle$ for different $V_{bias}$ at low doping. The energy on the vertical axis is relative to the electron affinity $\chi$ of bulk Si and the metal chemical potential $\mu_\mathrm{L}$. The vertical lines indicate $\phi_F$ at $V_{bias}$ = 0.02 V (blue, solid) and $V_{bias}$ = 0.2 V (red, dashed). (c) Solid curves: spectral current density $I(E)$ for different $V_{bias}$ at low doping. The dashed line indicates the value of  $\Phi^\mathrm{pot}$. $\langle V_H \rangle$ and $I(E)$ curves calculated at increasingly higher $V_{bias}$ are shown in blue$\to$green$\to$yellow$\to$red color scale. (d,e) Same as (b,c), but for intermediate doping.}
\label{fig:figure7}
\end{figure}

\begin{figure}
    \includegraphics[width=0.425\textwidth]{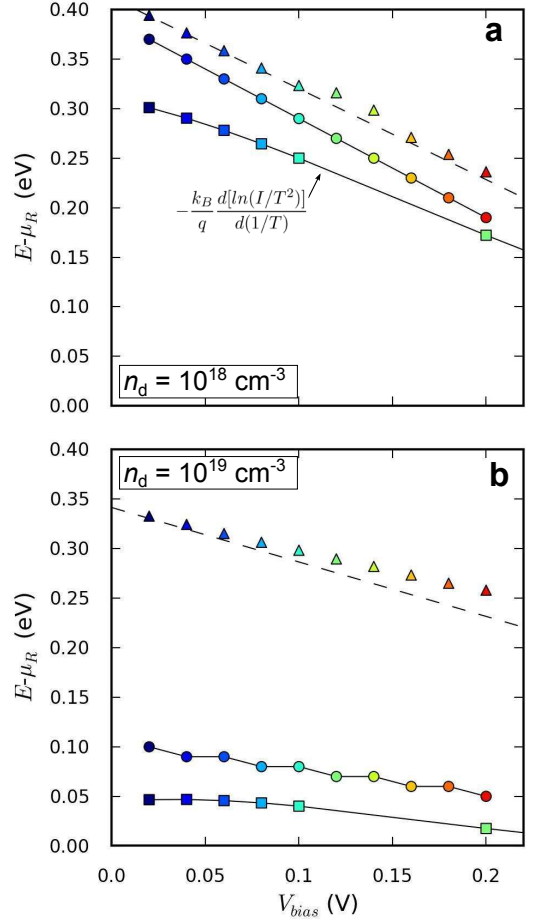}
    \caption{(a) Filled circles: energy of maximum spectral current $E(I^\mathrm{max})$ in Fig. \ref{fig:figure7}c as a function of $V_{bias}$ at low doping. The solid line is a guide to the eyes. Filled squares: variation of the slope-dependent term of Eq. \ref{eq:thermionic_3} (same as in Fig. \ref{fig:figure6}c). The solid line is a guide to the eyes. Filled triangles: $\phi_F$ as a function of $V_{bias}$. The dashed line shows the bias dependence $V_{bias}/n$ from Eq. \ref{eq:thermionic_3}. The energy on the vertical axis is relative to the semiconductor chemical potential $\mu_R$. (b) Same as (a), but for intermediate doping.}
\label{fig:figure11}
\end{figure}

A strong advantage of the DFT+NEGF simulations is that they allow the visualization of the electronic structure of the interface and the direct tracking of its changes when $n_\mathrm{d}$ and $V_{bias}$ are varied. This makes it possible to analyze the calculated $I$-$V_{bias}$ characteristics in terms of the electronic structure of the interface.

Fig. \ref{fig:figure4} shows the local density of states \cite{LDOS} (LDOS) of the two-probe model at equilibrium ({\em i.e.}, at $V_{bias}$ = 0 V) along the direction normal to the interface at the different doping densities considered. Increasing the doping has a two-fold effect on the electronic properties of the system: on the one side, $W_\mathrm{D}$ decreases from $\sim$200 nm to $\sim$20 nm when the doping is increased from $n_\mathrm{d}$ = 10$^{18}$ cm$^{-3}$ to $n_\mathrm{d}$ = 10$^{20}$ cm$^{-3}$, as a direct consequence of the increased screening of the n-doped silicon. Increasing $n_\mathrm{d}$ also shifts the Fermi level towards the silicon conduction bands. In particular, at $n_\mathrm{d}$ = 10$^{18}$ cm$^{-3}$ the conduction band minimum (CBM) of silicon at Z $>$ $W_\mathrm{D}$ lies at $E-\mu_{L,R}$ = +20 meV, whereas at $n_\mathrm{d}$ = 10$^{19}$ cm$^{-3}$ and $n_\mathrm{d}$ = 10$^{20}$ cm$^{-3}$ it lies at $E-\mu_{L,R}$ = --40 meV and $E-\mu_{L,R}$ = --100 meV, respectively. It is also worth noticing how the macroscopic average \cite{Baldereschi1988} of the Hartree potential along the direction normal to the interface, $\langle V_H \rangle$ (blue lines in Fig. \ref{fig:figure4}), follows the profile of the silicon CBM close to as well as far away from the interface. Similarly to what happens for the electronic bands, $\langle V_H \rangle$ becomes constant at Z $>$ $W_\mathrm{D}$, indicating that the electronic structure starts to resemble that of the infinite periodic bulk. A closer analysis also reveals that a finite density of states extends considerably on the semiconductor side of the interface, due to penetration of the metallic states into the semiconductor side \cite{Heine1965,Louie1975,Louie1976}.

Due to the lack of a well-defined electronic separation between the
metal and  the semiconductor, it is difficult to provide an
unambiguous value for $\Phi$ based on the electronic structure
data only. However, due to the fact that $\langle V_H \rangle$ closely traces the CBM, it is still possible to estimate the Schottky barrier by defining 
$\Phi^\mathrm{pot}$ as the difference between $\mu_{L}$ 
and the maximum of $\langle V_H \rangle$ on the semiconductor side of the interface, $\langle V_H^\mathrm{max}\rangle$ (see Fig. \ref{fig:figure4}).

We calculate $\Phi^\mathrm{pot}$ = 412 meV and $\Phi^\mathrm{pot}$
 = 342 meV for $n_\mathrm{d}$ = 10$^{18}$ cm$^{-3}$ and
 $n_\mathrm{d}$ = 10$^{19}$ cm$^{-3}$, respectively. For
 $n_\mathrm{d}$ = 10$^{20}$ cm$^{-3}$ the barrier is considerably lower, $\Phi^\mathrm{pot}$ = 133 meV, reflecting the more pronounced Ohmic behavior observed in the $I$-$V_{bias}$ curves. Focusing on the low and intermediate doping cases, it can be noticed how the values of $\Phi^\mathrm{pot}$ are considerably larger than those of $\Phi^\mathrm{AE}$ at V$_\mathrm{bias}$ $\to$ 0 V. In
 particular, at low doping $\Phi^\mathrm{pot} - \Phi^\mathrm{AE}$ = 112 meV, whereas at
 intermediate doping the difference is even larger, $\Phi^\mathrm{pot} - \Phi^\mathrm{AE}$ = 286 meV.

A consistent physical picture that rationalizes the $I$-$V_{bias}$ curves can be obtained by studying the doping dependence of the spectral current $I(E) = \frac{2q}{h} T(E,\mu_L,\mu_R) [f(\frac{E-\mu_L}{k_B T}) - f(\frac{E-\mu_R}{k_B T})]$. Fig. \ref{fig:figure7}b,d shows the profiles of $\langle V_H \rangle$ obtained at forward bias in the bias range 0.02 V $< V_{bias} <$ 0.2 V for low and intermediate doping densities. As $V_{bias}$ is increased, $\langle V_H^\mathrm{max} \rangle$ shifts towards higher energies due to image-force effects \cite{Sze}, and $\langle V_H \rangle$ becomes progressively flatter on the semiconductor side. The overall result of these changes is a decrease of the barrier $\phi_F$ associated with the thermionic emission process from the Si(100) conduction band to Ag(100) (see Fig. \ref{fig:figure7}a):
\begin{equation}
\phi_F = \Phi - V_{bias}/n.
\label{eq:phi_1}
\end{equation}
The associated spectral currents $I(E)$ are shown in Fig. \ref{fig:figure7}c,e. For an interface in which the only contribution to transport comes from thermionic emission, $I(E)$ should be non-zero only at $E-\mu_L > \Phi^\mathrm{pot}$. However, in the present case $I(E)$ is finite also at $E-\mu_L < \Phi^\mathrm{pot}$,  indicating that electron tunneling has a non-negligible contribution to $I$. This contribution is much larger for intermediate than for low doping densities. Indeed, at $V_{bias} \to$ 0 V, the position of $E(I^\mathrm{max})$ lies very close to $\Phi^\mathrm{pot}$ in the low doping case, as expected in the case of a nearly ideal Schottky diode. Conversely, at intermediate doping $E(I^\mathrm{max})$ lies well below $\Phi^\mathrm{pot}$, indicating that electron tunneling has become the dominant transport process.

The trend of $I(E)$ with $V_{bias}$ is consistent with these considerations. At low doping, $E(I^\mathrm{max})$ is pinned to $\langle V_H^\mathrm{max} \rangle$, whereas the onset of finite $I(E)$ at $E-\mu_L < \Phi^\mathrm{pot}$ moves towards higher energies, following the variation of $\langle V_H \rangle$. On the other hand, at intermediate doping the overall shape of $I(E)$ remains the same as $V_{bias}$ is increased, and the variation of $E(I^\mathrm{max})$ follows closely that of $\langle V_H \rangle$. We also notice the presence of a narrow resonance at $E-\mu_L$ = +0.395 eV, whose position is independent on $n_\mathrm{d}$ and $V_{bias}$. This is due to a localized electronic state at the interface which is pinned to $\mu_L$.

The variation of $\phi_F$ with $V_{bias}$ can be related to the slope-dependent term of Eq. \ref{eq:thermionic_3} by assuming $\Phi$  = $\Phi^\mathrm{AE}$ in Eqs. \ref{eq:thermionic_3}-\ref{eq:phi_1}, thus allowing for a direct comparison with the AE data (see Fig. \ref{fig:figure11}). Independently of the value of $V_{bias}$, the slope-dependent term lies always below $\phi_F$, due to the missing contribution of electron tunneling in the AE method: the latter assumes that the current has a purely thermionic origin, and consequently predicts a value of $\phi_F$ lower than the actual one. In agreement with the previous analyses, this deviation is considerably larger in the intermediate doping case, due to the much larger contribution of electron tunneling.

\subsection{\label{sec:Comparison of the two-probe with the slab model}Comparison of the two-probe with the slab model}

\begin{figure}
    \includegraphics[width=0.4\textwidth]{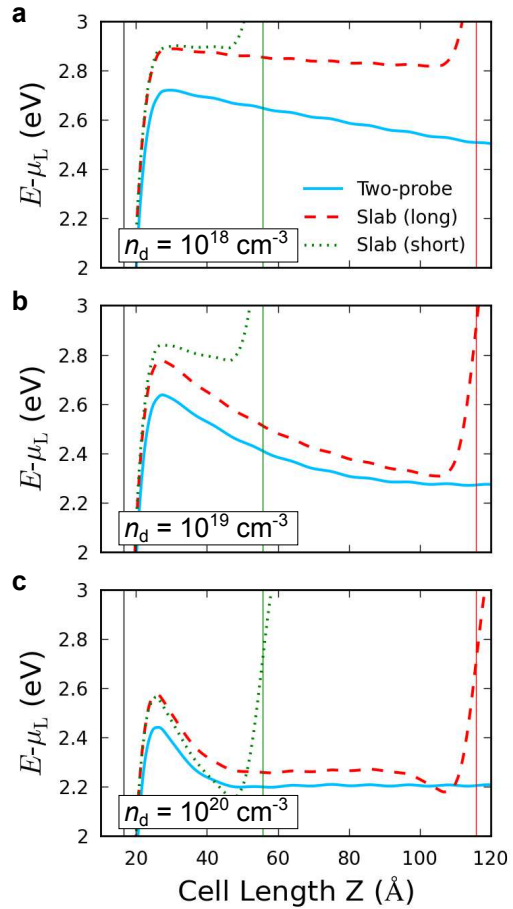}
    \caption{Profile of $\langle V_H \rangle$ along the direction Z normal to the interface plane, calculated for the two-probe setup (blue solid line) and for the short (green dotted line) and long (red dashed line) slab models, at doping densities $n_\mathrm{d}$ = 10$^{18}$ cm$^{-3}$ (a), $n_\mathrm{d}$ = 10$^{19}$ cm$^{-3}$ (b) and $n_\mathrm{d}$ = 10$^{20}$ cm$^{-3}$ (c). The vertical black solid line indicated the position of the interface. The vertical green (red) line indicates the position Si(100) layer farthest from the interface in the short (long) slab model.}
\label{fig:figure8}
\end{figure}

\begin{figure}
    \includegraphics[width=0.425\textwidth]{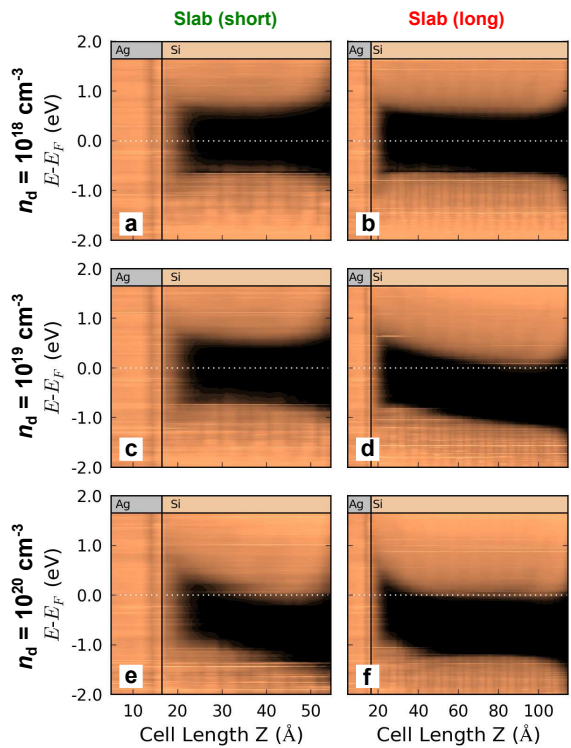}
    \caption{LDOS of the short (a,c,e) and long (b,d,f) slab models for different effective doping densities. Doping densities: (a,b) $n_\mathrm{d}$ = 10$^{18}$ cm$^{-3}$; (c,d) $n_\mathrm{d}$ = 10$^{19}$ cm$^{-3}$; (e,f) $n_\mathrm{d}$ = 10$^{20}$ cm$^{-3}$. The energy on the Y-axis is scaled with respect to the system Fermi energy $\mathrm{E_F}$.}
\label{fig:figure9}
\end{figure}

The results obtained using the two-probe model can be used as a
reference to validate the use of finite-size models to describe the
Ag(100)/Si(100) interface. Such models are integral parts of the
band alignment method often used to evaluate $\Phi$
using conventional DFT
\cite{VandeWalle1987,VandeWalle1989,Fraciosi1996,Peressi1998}. The
method relies on aligning the electronic band structures of the two
bulk materials forming the interface on an absolute energy scale by
using a reference quantity, often $\langle V_H \rangle$
\cite{Baldereschi1988}. The perturbation of the bulk electronic
structure in each material due to the presence of the interface is
accounted for by either a slab \cite{Butler2011} or a fully periodic
\cite{Niranjan2006} model. $\langle V_H \rangle$ is then used as a common reference to align the electronic structure obtained from independent calculations of the two bulk materials. Despite its widespread use, this strategy relies on two drastic assumptions. Firstly, it is implicitly assumed that the electronic properties of the interface are independent of the doping level of the semiconductor. Moreover, it is assumed that the electronic properties in the central part of each side of the interface model are a good approximation to those of the two bulk materials.

Fig. \ref{fig:figure8} shows a comparison between  $\langle V_H \rangle$ obtained at the different doping densities considered for the two slab models (short and long, see Section \ref{sec:System}) and for the two-probe setup. We notice that introducing an effective doping in the slab model, which was not taken into account in previous slab models for Ag/Si interfaces \cite{Butler2011}, attempts to better mimic the two-probe simulation in which silver is interfaced with n-doped silicon. The profiles of $\langle V_H \rangle$ of the three different systems have been aligned according to the value of $\mu_L$, the side at which Dirichlet boundary conditions are used for the three systems. Irrespectively of the doping level, the doped short slab model provides a poor description of the variation of $\langle V_H \rangle$ at the interface. In particular, $\langle V_H^\mathrm{max} \rangle$ is always $\sim$200 meV higher that that obtained for the two-probe model. Furthermore, on the Si(100) side of the interface, $\langle V_H \rangle$ does not decay correctly with the distance from the interface for the short slab model and, even more importantly, it does not converge to a constant value. The situation improves by increasing the width of the Si(100) layer. For the long slab model, at increasingly larger doping densities the profile of  $\langle V_H \rangle$ resembles more and more that of the two-probe model. Indeed, in the best case scenario, \emph{i.e.}, at $n_\mathrm{d} = $10$^{20}$ cm$^{-3}$, the profile of $\langle V_H \rangle$ evaluated using the long slab becomes constant in the center of the Si(100) region, albeit  still higher than that of the reference by $\sim$100 meV. The limitations of the slab model in reproducing the electronic structure at the interface are also evident by looking at the corresponding LDOS plots (see Fig. \ref{fig:figure9}). Similarly to what is observed for $\langle V_H \rangle$, the short slab model fails to reproduce the band bending observed at low doping, as well as the correct trend in the decrease of W$_\mathrm{D}$ as doping is increased. The latter is qualitatively reproduced using the long slab model. However, these modest improvements going from the short to the long slab model come at the expenses of a much higher computational cost. In fact, each DFT calculation for the long slab model takes on average 338.6 s/step. Conversely, each DFT+NEGF calculation using the two-probe model is approximatively one order of magnitude faster, taking on average 46.6 s/step. This suggests that, in addition to computational efficiency, there are also more fundamental reasons for making DFT+NEGF the method of choice for describing  M-S interfaces, as in the two-probe setup the two main assumptions of the band alignment method are naturally lifted. We emphasize that, although the results presented in this paper are specific to the Ag(100)/Si(100) interface only, similar conclusions are likely to hold true for all systems for which the poor screening on the semiconductor side of the interface results in space--charge effects that extend over widths of the order of several nanometers.

\section{\label{sec:Conclusions} Conclusions}

In this work, we have presented an approach based on density functional theory (DFT) and non-equilibrium Green's functions (NEGF) for realistic metal-semiconductor (M-S) interfaces modeling. Our  approach is designed to deal effectively and correctly with the non-periodic nature of the interface, with the semiconductor band gap and with the doping on the semiconductor side of the contact, and allows for a direct theory-experiment comparison  as it can simulate $I$-$V_{bias}$ characteristics. Using a Ag/Si interface relevant for photovoltaic applications as a model system, we have shown that our approach is a better alternative to (i) analytical approaches such as the ``Activation Energy" (AE) method to analyze the $I$-$V_{bias}$ characteristics of non-ideal rectifying system with non-negligible tunneling contribution, and (ii) finite-size slab models to describe the interface between metals and doped semiconductors. This DFT+NEGF approach could pave the way for a novel understanding of M-S interfaces beyond the limitations imposed by traditional analytical and atomistic methods.

\section*{\label{sec:Acknowledgements} Acknowledgements}
QuantumWise acknowledges support from Innovation Fund Denmark, grant Nano-Scale Design Tools for the Semi-conductor Industry (Grant No. 79-2013-1). The research leading to these results has received funding from the European Community’s Seventh Framework Programme (FP7/2007-2013) under grant agreement III-V-MOS Project n619326). We thank Walter A. Harrison, Andreas Goebel, and Paul A. Clifton at Acorn Technologies for their input on this work. DS acknowledges support from the H.C. {\O}rsted-COFUND postdoc program at DTU.

\bibliography{biblio}

\begin{thebibliography}{57}
\bibitem [{LDO()}]{LDOS}%
  \BibitemOpen
  \href@noop {} {}\bibinfo {note} {The LDOS is obtained by projecting the
  density of states on each atomic layer in the Z direction and then
  interpolating the layer-resolved density of states to obtain a continuous
  LDOS. An exponential scale is used for the LDOS intensity. While this
  procedure provides a good representation of the silicon band gap throughout
  the device, the finer details of the LDOS plots depend on (i) the
  interpolation procedure used to obtain a continuous LDOS and (ii) the
  exponential scale used in the interpolated LDOS plots.}\BibitemShut {Stop}%
\end{thebibliography}

\end{document}